\documentclass[preprint,3p,times,twocolumn,authoryear]{elsarticle}
\usepackage{graphicx}
\usepackage{amsmath,amssymb}
\usepackage{hyperref}
\usepackage{derivative}
\usepackage{lineno}
\usepackage{soul, color}

\DeclareUnicodeCharacter{2212}{-}

\begin{document}

\begin{frontmatter}

\title{Variability study of classical supergiant X-ray binary 4U 1907+09 using \textit{NuSTAR}}

\author[label1,label2]{Raj Kumar\corref{cor1}}
\ead{arya95raj@gmail.com}
\author[label3]{Sayantan Bhattacharya}
\author[label3]{Sudip Bhattacharyya}
\author[label1,label2]{Subir Bhattacharyya}

\cortext[cor1]{Corresponding author}
\affiliation[label1]{organization={Astrophysical Sciences Division, Bhabha Atomic Research Centre},
             city={Mumbai},
             postcode={400085},
             country={India}}

\affiliation[label2]{organization={Homi Bhabha National Institute},
             city={Mumbai},
             postcode={400094},
             country={India}}
\affiliation[label3]{organization={Department of Astronomy and Astrophysics, Tata Institute of Fundamental Research},
             addressline={1 Homi Bhabha Road, Colaba},
             city={Mumbai},
             postcode={400005},
             country={India}}

\begin{abstract}
We investigate the X-ray variability of the supergiant X-ray binary 4U 1907+09 using the new \textit{NuSTAR} observation of 2024. Unlike the previous \textit{NuSTAR} observations, the source shows significant flux variation during the current one. The light curve exhibits dips (off-state) and flares (on-state). The phase-coherent timing analysis during the on-state yields a pulse period of $443.99(4)~\mathrm{s}$, showing the pulsar's continued spin-down. The pulse profiles show an asymmetric double-peaked structure with a phase separation of 0.47 between the two peaks. A cyclotron resonance scattering feature (CRSF) is also detected at $\sim 17.6~\mathrm{keV}$, along with its harmonic at $\sim 38~\mathrm{keV}$, persisting across all flux states. Flux-resolved spectroscopy reveals that the CRSF energy remains constant despite a 25-fold change in flux. The spectral parameters, like photon index and e-fold energy, are out of phase with the pulse shape, whereas the cutoff energy is in phase with the pulse shape. The source's luminosity during the on-state is $2.85 \times 10^{35}~\mathrm{erg~s^{-1}}$, consistent with a ``pencil'' beam radiation pattern expected at this flux level from a collisionless gas-mediated shock. These results offer further insights into the accretion dynamics and magnetic field geometry of this system.
\end{abstract}


\end{frontmatter}


\section{Introduction} 
\label{sec:intro}
Supergiant X-ray binaries (sgXBs) are binary systems that consist of a compact object and a massive O/B supergiant star. In most cases, the accretion of stellar wind from the supergiant star onto a compact object is responsible for the X-ray emission in these systems. There are two types of X-ray binaries: classical supergiant high-mass X-ray binaries (HMXRBs) and supergiant fast X-ray transients (SFXTs) \citep{Pradhan2018}. 
The classical supergiant HMXRBs are persistent, with luminosity above $\sim10^{35}\;\mathrm{erg\;s^{-1}}$. These systems were seen to vary by a factor of $10-100$ \citep{Walter2015, Kretschmar2019}. SFXTs have low luminosity ($10^{33}-10^{34}\;\mathrm{erg\;s^{-1}}$) and produce short and strong flares with typical lifetimes of hours, reaching luminosity $10^{35}-10^{37}\;\mathrm{erg\;s^{-1}}$ \citep{Sidoli2018, Kretschmar2019}.

4U 1907+09 is a classical supergiant high-mass X-ray binary identified in the third \textit{Uhuru} survey \citep{Giacconi1971}. The pulsar orbiting companion with an eccentricity $e\sim0.28$ and an orbital period of $P\sim8.3753$ d \citep{intzand1998}. Optical and infrared studies suggest an O8-O9 Ia type supergiant donor \citep{cox2005, Nespoli2008}. The literature reports a source distance of 5 kpc \citep{cox2005, Nespoli2008}. However, the distance to the source calculated by Gaia EDR3 is around 1.9 kpc \citep{Bailer2021}.
X-ray pulsation with a period of 437.5 s was initially reported by \cite{Makishima1984} utilizing \textit{Tenma} observations. 
Early observations of the source revealed a relatively constant spin-down rate \citep{Baykal2006}; however, further observations revealed several torque reversals and spin-down rate variations \citep{Fritz2006, Inam2009}. \textit{INTEGRAL} and \textit{RXTE} measurements show short-term fluctuations in the pulse period compared to long-term changes in spin period rates, consistent with the random walk model \citep{Sahiner2012}. The most recent studies reported spin down of the pulsar \citep{Varun2019, Tobrej2023}.

4U 1907+09 is a variable X-ray source that exhibits irregular flaring and dipping activities. During the dips, \cite{Intzand1997} did not detect X-ray pulsed emission from 4U 1907+09 using \textit{RXTE} observations. The typical duration of the dips ranges from a few minutes to 1.5 hours. They suggested that cessation of accretion from the inhomogeneous wind of the companion star may be responsible for dips. \cite{Doroshenko2012} investigated the dipping activity using \textit{Suzaku} observations and found the pulsations during the off state. \cite{Bozzo2008} reviewed the theory of wind accretion in HMXRBs and discussed it in the context of SFXT. Moreover, this model was employed by \cite{Doroshenko2012} to explain the transition between off-state (dips) and on-state (normal flux state) in 4U 1907+09. \cite{Shakura2012} proposed a theoretical model for quasi-spherical subsonic accretion onto slowly rotating pulsars. The accreting matter forms a hot quasi-static shell above the magnetosphere. This model was further used to explain the off states in Vela X-1, 4U 1907+09, and GX 301-2 \citep{Shakura2013}.

The X-ray spectra of 4U 1907+09 are characterized by a power law 
and an exponential cut-off 
\cite{Schwartz1980, Marshall1980, Makishima1984, Cook1987, Chitnis1993, Roberts2001, Coburn2002, Baykal2006, Fritz2006, Varun2019, Tobrej2023}. During the binary orbit, inhomogeneous accretion via intense stellar wind caused highly variable hydrogen column density ($n_H$) ranging from $1\times10^{22}$ to $9\times10^{22}$ cm$^{−2}$ \cite{Intzand1997}. Moreover, the spectra showed a narrow spectral line at around 6.4 keV, which is the Fe $\mathrm{K}\alpha$ fluorescence emission caused by the X-ray reflection of primary X-ray emission from the accretion disk or stellar wind.
Cyclotron-resonant scattering features (CRSFs) at $\sim19$ keV with a second harmonic at $\sim39$ keV have been seen using \textit{Ginga} \citep{Mihara1995PhDT} and \textit{BeppoSAX} observations \citep{Cusumano1998}. This implies a surface magnetic field strength of $2.1\times10^{12}$ G \citep{Cusumano1998}.
The flaring activity observed by \textit{AstroSat} indicates that the CRSF parameters have not been altered \citep{Varun2019}. 
The existence of cyclotron lines at $\sim18$ keV and $\sim38$ keV was recently verified by \cite{Tobrej2023} utilizing \textit{NuSTAR}. Furthermore, an absorption line possibly due to Ni $\mathrm{K}\alpha$ and Ni $\mathrm{K}\beta$ at $\sim8$ keV was also noticed.

In this work, we have estimated the spin rate and examined the flux-resolved spectral analysis of 4U 1907+09 using \textit{NuSTAR} data observed on 20 November 2024. The manuscript is organized as follows: Section \ref{sec:data} describes observation and data reduction. Section \ref{sec:results} presents the timing and spectral data analysis results. Section \ref{sec:discussion} discusses the results. Section \ref{sec:summary} summarises our findings.

\section{Observation and data reduction} \label{sec:data}
\textit{NuSTAR} is a NASA space mission that observes the sky in the hard X-ray energy band (3-79 keV) of the electromagnetic spectrum \citep{Harrison2013}. It comprises two co-aligned grazing incidence telescopes with specially coated optics and detector units (FPMA and FPMB). In this work, we analyzed the \textit{NuSTAR} data of 4U 1907+09 observed on MJD 60634. 

\begin{figure}[h]
    \centering
    \includegraphics[width=0.5\textwidth]{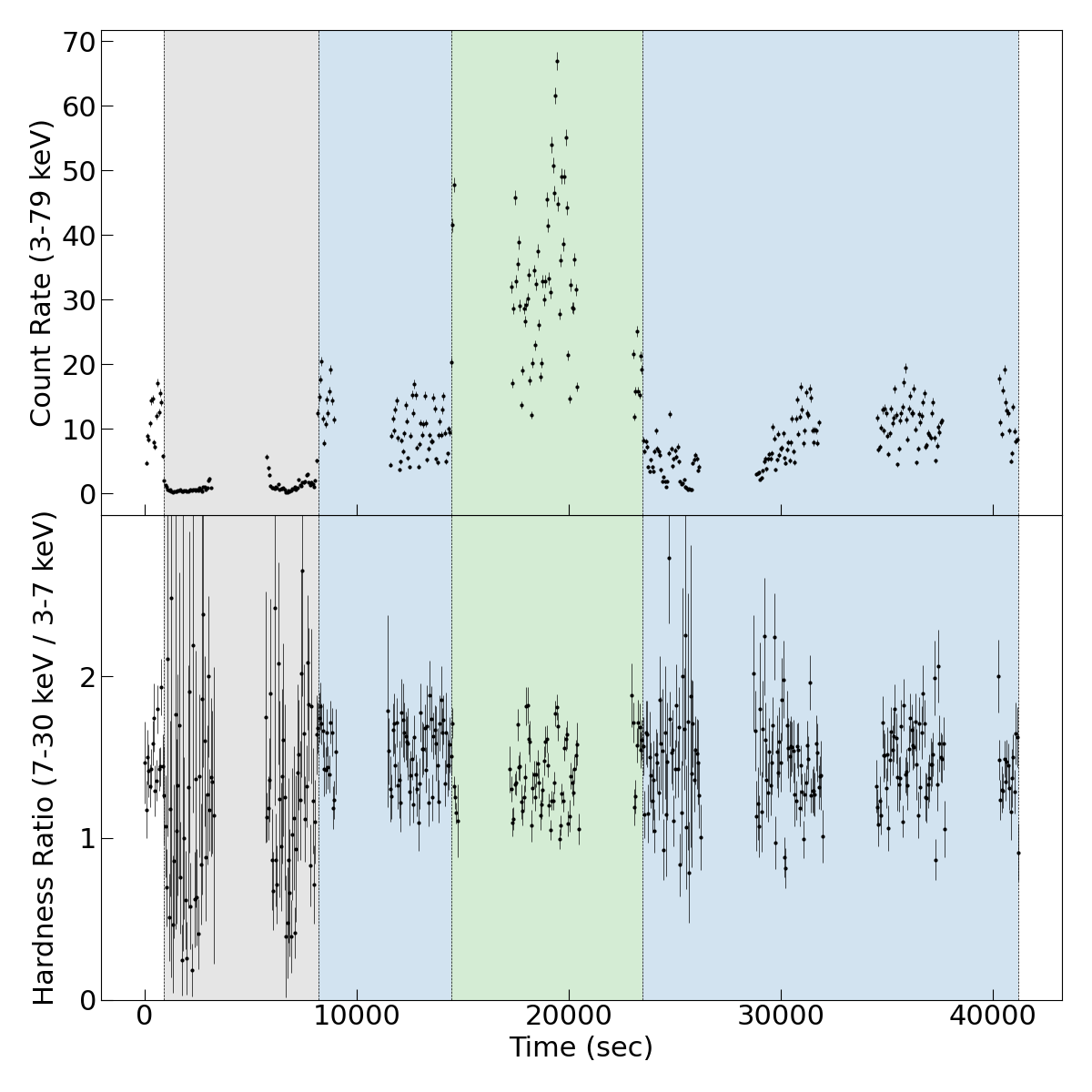}

    \caption{\textit{NuSTAR} light curve (3-79 keV) and hardness ratio (7-30 keV/ 3-7 keV) of 4U 1907+09 for FPMA with a time bin size of 60 sec is shown in upper and lower panels, respectively. The grey-shaded region shows the ``off-state". The blue and green-shaded region shows the ``on-state" with low and high flux, respectively. The data corresponding to the unshaded initial portion of lightcurve (~900 sec) is not used in this work.  (see section ~\ref{sec:results})}
    \label{nu_lc_fpma}
\end{figure}

\texttt{HEASOFT} version 6.33 and \texttt{NuSTARDAS V2.1.2} with \texttt{CALDBVER 20241126} were used to process and filter the \textit{NuSTAR} event data. 
We selected a 90 arcsec radius region around the source centroid for the source and a 90 arcsec radius region away from the source for background estimation. The science products: spectra, light curves, and response files were extracted using the \texttt{nuproducts} script for both modules (FPMA and FPMB). We extracted the light curves with a time resolution of 0.1 s in the 3-79 keV energy range. The \textit{NuSTAR}/FPMA light curve in the 3-79 keV energy range is shown in Figure \ref{nu_lc_fpma}. The source is largely variable during the observation. We divided the data into off-state and on-state. We extracted spectra and response files for the on and off states. We further divided the on-state into low and high luminosity and extracted the spectra and response files. The \texttt{grppha} task was used to rebin the source spectra for a minimum of 30 counts per bin.

\begin{figure}[h]
    \centering
    \includegraphics[width=0.5\textwidth]{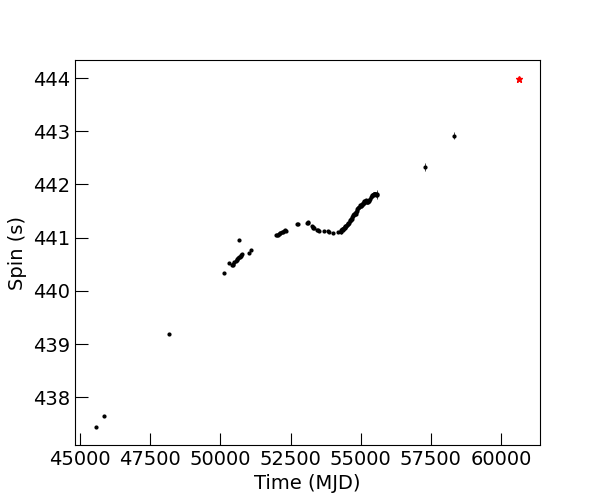}
    \caption{Pulse period history of 4U 1907+09. The pulse period measurements are take from \cite{Cook1987, intzand1998, Baykal2001, Baykal2006, Mukerjee2001, Fritz2006, Inam2009, Sahiner2012, Varun2019, Tobrej2023}. The red star corresponds to the spin period estimated in this work. (see section~\ref{subsec:timing})}
    \label{PHist}
\end{figure}

\begin{figure*}[h]
\centering
\begin{minipage}{0.3\textwidth}

\includegraphics[angle=0,scale=0.4]{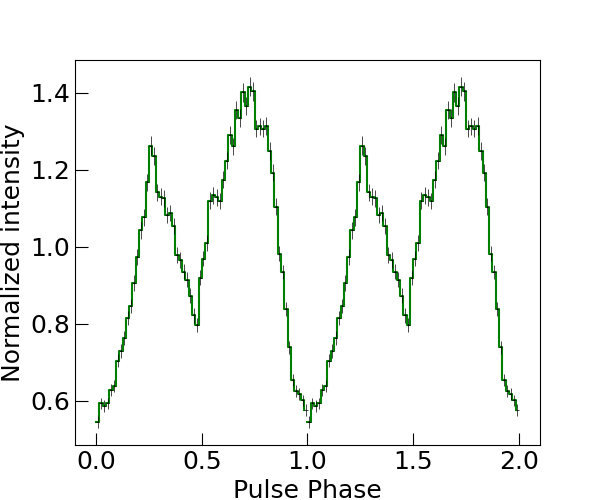}
\end{minipage}
\hspace{0.15\linewidth}
\begin{minipage}{0.3\textwidth}
\includegraphics[angle=0,scale=0.4]{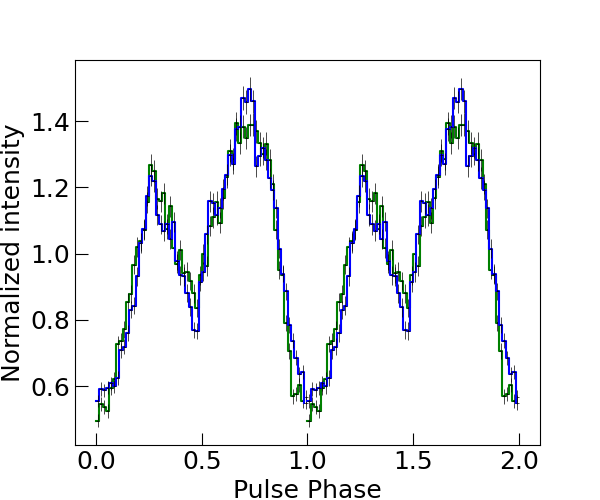}
\end{minipage}
\hspace{0.15\linewidth}

\caption{Left panel: Pulse profile of 4U 1907+09 in 3-79 keV energy range using FPMA. Right panel: The pulse profile of 4U 1907+09 in high and low flux on-state. The green color corresponds to the low flux state, and the blue color corresponds to the high flux state. (sec section~\ref{subsec:timing})}
\label{pulse}
\end{figure*}

We extracted the light curves during on state with a time resolution of 0.1 s in the following energy ranges: 3–10 keV, 10–15 keV, 15–20 keV, 20–30 keV, 30–50 keV, and 3–79 keV. We used \texttt{barycorr} tool to correct the photon arrival times to the barycenter of the solar system using the DE-430 ephemeris \citep{folkner2014planetary}. 

\section{Results} \label{sec:results}
During this \textit{NuSTAR} observation, the source was found to have different X-ray fluxes. We divided the data into off-state and on-state. The effective on-state and off-state exposures are 4.5 ks and 16.5 ks, respectively. Further, we divided the on state into high-flux and low-flux states. The effective exposure during high flux and low flux in the on state are 4.1 ks and 12.4 ks, respectively. The light curve, Hardness ratio and different states are shown in Figure \ref{nu_lc_fpma}. First, we performed temporal and spectral analyses on the averaged on-state. 

\subsection{Timing analysis} \label{subsec:timing}
\begin{figure}[]
\centering
\includegraphics[angle=0,scale=0.7]{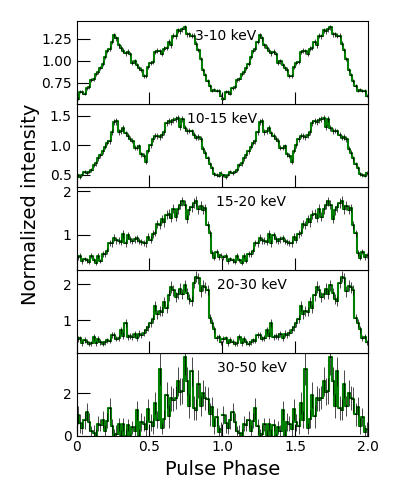}
\caption{Energy-dependent pulse profiles of 4U 1907+09 using FPMA. The energy range for the pulse profiles is specified inside the panels. (see section~\ref{subsec:timing})}\label{Epulse}
\end{figure}

\begin{figure}[]
    \centering
    \includegraphics[width=0.5\textwidth]{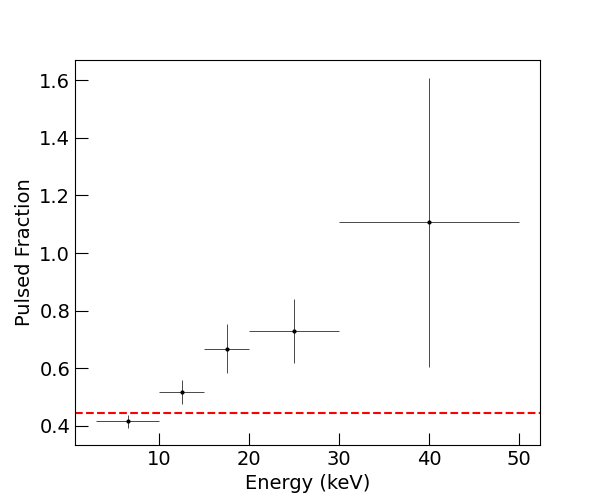}
    \caption{The energy dependence of pulsed fraction. The dashed red line corresponds to the pulse fraction in the 3-79 keV energy range. (see section~\ref{subsec:timing})}
    \label{PF}
\end{figure}

We examined the 3-79 keV light curve of averaged on-state for coherent pulsations from the source. We employed the \texttt{xronos} tool \texttt{efsearch} to estimate the best rotational period. The best period measured in the 3-79 keV energy range is 443.99(4) (MJD 60634.05). The uncertainty in the period was estimated using the bootstrap method by simulating 1000 light curves \citep{Boldin2013}. The calculated pulse period value with pulse period history of 4U 1907+09 is shown in Figure \ref{PHist}. The source was found to be in a spin-down phase.
We employed the \texttt{efold} technique \citep{Leahy1983} to create folded light curves at the best period in 3-79 keV, 3-10 keV, 10-15 keV, 15-20 keV, 20-30 keV and 30-50 keV. The left panel of Figure \ref{pulse} shows the pulse profile in the 3-79 keV energy range. Figure \ref{Epulse} depicts the pulse shape in different energy ranges. There is a noticeable energy dependence in the pulse profile.
A double-peaked feature is visible in the 3-79 keV energy range. This feature dominates in the 3-10 and 10-15 keV energy range. The peak at $\sim$ 0.25 starts to fade after the 15 keV. 
The pulse fraction (PF) is calculated using the expression,
$$\mathrm{PF} =  \frac{(I_\mathrm{max} − I_\mathrm{min} )}{(I_\mathrm{max} + I_\mathrm{min} )}$$
Where $I_\mathrm{max}$ is the maximum normalized luminosity and $I_\mathrm{min}$ is the minimum normalized luminosity. The energy-dependent PF is shown in Figure \ref{PF}.
We checked the flux-dependent pulse shape as shown in the right panel of Figure \ref{pulse}. The pulse shape is similar in both low and high flux on-states.

\subsection{Spectral analysis} \label{subsec:spectral}
\subsubsection{Averaged on state spectral analysis} 
We used \texttt{XSPEC V12.14.0h} \citep{Arnaud1996} for spectral modeling of the data. A power-law model, along with an energy cutoff and smoothing function, is used to represent the continuum spectra observed in X-ray pulsars. We fitted the data by model: \texttt{const*tbabs(highecut*pow)}. 
The model \texttt{const} cross-calibrates instruments FPMA and FPMB. \texttt{tbabs} is the Tuebingen-Boulder interstellar medium absorption model.
\texttt{pow} is the photon power-law model. \texttt{highecut} is the high-energy cutoff model given by
\begin{equation*}
\displaystyle M(E) = \begin{array}{ll} exp\left[(E_c-E)/E_f\right] & E \geq E_c\\ [.2cm]1.0 & E \leq E_c \end{array}
\end{equation*}
where $E_c$ is cutoff energy (keV) and $E_f$ is e-folding energy (keV). An artificial discontinuity can be observed at the cutoff energy, and in those cases it is smoothed by including a Gaussian absorption line tied to the cutoff energy. We checked for this feature in our observations and, due to the absence of any aforementioned discontinuity, we did not use the additional Gaussian component during the fitting. The data does not effectively constrain the column density. So, we fixed the $n_H$ value at $1.5\times10^{22}\;\mathrm{cm^{-2}}$ \citep{Maitra2013, Varun2019}. The $\chi^2/\mathrm{dof}$ came out to be $\sim2202/1119$. An iron line and absorption features around 18 keV and 35 keV due to cyclotron absorption are visible in the residual.
So, we added model \texttt{gauss} for the Fe feature and a multiplicative model \texttt{gabs}. The model becomes \texttt{const*tbabs(gauss+highecut*pow)*gabs*gabs}. The $\chi^2/\mathrm{dof}$ improved to $\sim1277/1110$. The best fit yielded $\Gamma\sim1.12$ and $E_\mathrm{cutoff}\sim11.8$ keV, and $E_\mathrm{fold}\sim12.8$ keV.
The Gaussian line is fitted at 6.31 keV with an equivalent width equal to 46 eV, and the absorption lines came out at 17.6 keV and 38.02 keV. The flux came out to be $6.59_{-0.04}^{+0.04}\times10^{-10}\;\mathrm{erg/cm^2/s}$.\\
To assess the importance of the absorption features in the spectra, we used the \texttt{simftest} script from \texttt{xspec}. It uses Monte Carlo simulations to produce simulated data sets from the original data, which are then used to estimate $\Delta\chi^2$ for \texttt{gabs} components. We conducted $10^4$ simulations for each \texttt{gabs} component.
The difference between observed $\chi^2$ and maximum $\chi^2$ obtained from simulated data sets for $\sim17.6$ keV and $\sim38.02$ keV \texttt{gabs} component are $\sim400$ and $\sim54$. It confirms the $>5\sigma$ detection of absorption features at $\sim17.6$ keV and $\sim38.02$ keV.\\

Then we fitted the data by model \texttt{const*tbabs(gauss+compTT)*gabs*gabs}. \texttt{compTT} \citep{Titarchuk1994} is an analytic model describing the Comptonization of soft photons in hot plasma and is suitable for describing the continuum shape from X-ray pulsars. The seed soft photon temperature, plasma temperature, and plasma optical depth are $\sim0.52$ keV, $\sim7.87$ keV, and $\sim6.26$, respectively. The $\chi^2/\mathrm{dof}$ for this model is $\sim1285/1110$, which is slightly higher than the previous model.\\

\begin{figure}[]
    \centering
    \includegraphics[width=0.5\textwidth, angle=0]{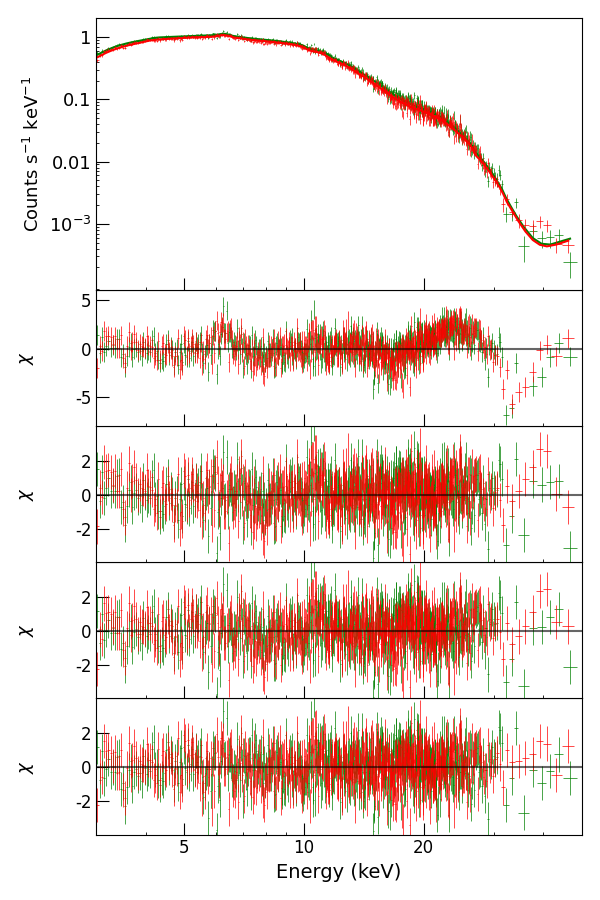}
    \caption{Top: The \textit{NuSTAR} spectra of 4U 1907+09. The green and red colour corresponds to FPMA and FPMB data, respectively. Second panel: residual of the continuum only model-\texttt{const*tbabs(highecut*pow)}. The emission line at 6.5 keV and absorption features at 18 keV and 35 keV are visible. Third panel: residual of the model-\texttt{const*tbabs(gauss+highecut*pow)gabs*gabs}. Fourth panel: residual of the model-\texttt{const*tbabs(gauss++compTT)gabs*gabs}. Bottom panel: residual of the model-\texttt{const*tbabs(gauss+bbodyrad+compmag)gabs*gabs}. (see section~\ref{subsec:spectral})}
    \label{spectra}
\end{figure}

Finally, we employed the \texttt{compmag} \citep{Farinelli2012} model to fit the continuum spectra. The \texttt{compmag} model describes the spectral formation in the accretion column onto the polar cap of a magnetized neutron star, taking into account both thermal and bulk Comptonization processes. Since the magnetic field of 4U 1907+09 is $2\times10^{12}$ G, the \texttt{compmag} model is suitable for this system. The model becomes \texttt{const*tbabs(gauss+bbodyrad+compmag)*gabs*gabs}. During the fitting, we consider that the accretion velocity increases towards the NS (by fixing \texttt{betaflag}=1) with the index of velocity profile $\eta = 0.5$ and terminal velocity at the NS surface $\beta_0 = 0.05$. We fixed the accretion column $r_0 = 0.25$ (in units of the NS Schwarzschild radius) and NS albedo A = 1. 
Since the norm of \texttt{compmag} is given by $R^2_{km}/D^2_{10}$, where $R_{km}$ is the radius of the accretion column in km and $D_{10}$ is the distance of the source in units of 10 kpc. Considering the radius of the accretion column is 1 km, and the source distance is equal to 1.9 kpc. We fixed the norm at 28. The temperature of the seed blackbody temperature ($kT_{bb}$) of \texttt{compmag} is tied with the blackbody temperature of \texttt{bbodyrad}. The remaining parameters of this spectral model were kept free. The $\chi^2/\mathrm{dof}$ for this model is $1236/1110$. The optical depth of the accretion column came out to be $\sim1.75$. Since the \texttt{compmag} model is applicable for $\tau < 1$, we did not further use this model. Also, the low luminosity of the source disfavors the formation of the accretion column. The spectral parameters for different models used in this work are given in Table \ref{table_par}. The \textit{NuSTAR} spectra with residuals of different models used in this work are shown in Figure \ref{spectra}.

\subsubsection{Phase resolved Spectroscopy}
\label{sssec:phase}
\begin{figure}[]
    \centering
    \includegraphics[width=0.5\textwidth]{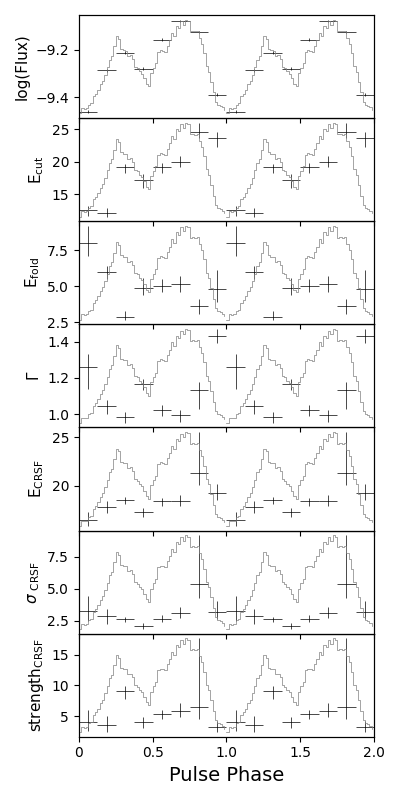}
    \caption{Phase resolved spectroscopy of 4U 1907+09. The flux is calculated in the 3-25 keV energy range. (see section~\ref{sssec:phase})}
    \label{phase_res}
\end{figure}

We extracted the spectral files into eight phases: 0-0.125, 0.125-0.25, 0.25-0.375, 0.375-0.5, 0.5-0.625, 0.625-0.75, 0.75-875, and 0.875-1.0. We fitted the spectral data with the model: \texttt{const*tbabs*(highecut*pow)gabs}. The variation of different spectral parameters with phase is given in Figure \ref{phase_res}. The top panel provides the 3.0-25.0 keV flux at various pulse phases, facilitating comparison with the pulse profile. The continuum fit parameters exhibit substantial fluctuation with the pulse phase. The cutoff energy and CRSF line center are correlated (in phase) with the pulse shape, whereas the e-fold energy and the photon index are out of phase with the pulse profile.

\subsubsection{Flux resolved spectral analysis} \label{sssec:flux}
\begin{figure}[]
    \centering
    \includegraphics[width=0.5\textwidth]{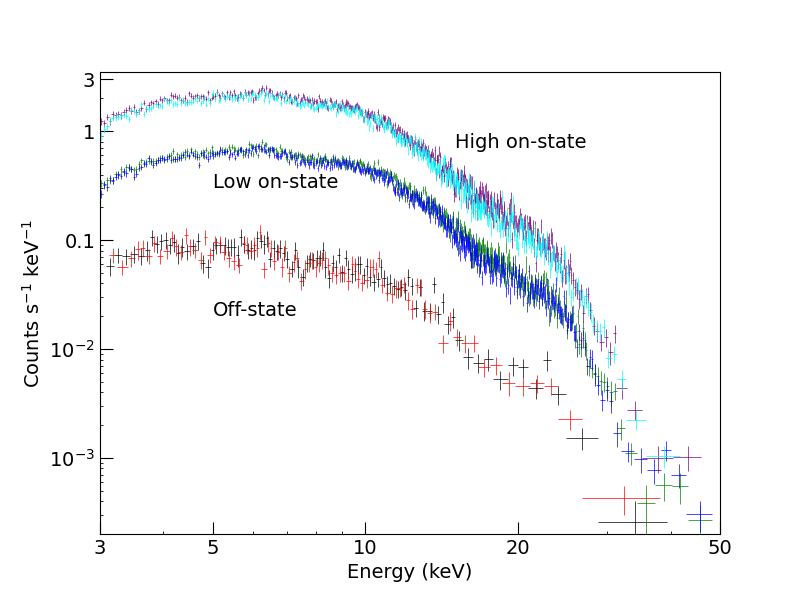}
    \caption{The spectral data of 4U 1907+09 in off-state, low and high flux on-state. (see section~\ref{sssec:flux}) }
    \label{nu_lum}
\end{figure}

As shown in Figure \ref{nu_lc_fpma}, 4U 1907+09 shows the flux variation in the light curve. The spectral data for the off-state, low-flux on-state, and high-flux on-state is shown in Figure \ref{nu_lum}. We fitted the off-state and high-flux on-state data with the model \texttt{const*tbabs(highecut*pow)*gabs}.
The low flux on-state data required additional models \texttt{gabs} and \texttt{gauss}. We fitted the low flux on-state data with the model.
\texttt{const*tbabs(gauss+highecut*pow)*\\gabs*gabs}. The $\Gamma$ for off-state, high flux on-state, and low flux on-state come out to be $\sim1.43$, $\sim1.12$, and $\sim1.08$, respectively. At various flux levels, the cyclotron line is found at $\sim18$ keV. The line width of CRSF during high flux on-state and low flux on-state is $\sim3.07$ and $\sim2.5$, respectively. Compared to low flux on-state, flux increased 3 times during high flux on-state, and flux decreased 8 times during off-state. The e-folding energy during the high flux on-state and low flux on-state is $\sim4.6$ and $\sim6.27$, respectively. The spectral parameters for each data are given in Table \ref{table_flux}. 

\begin{table*}[]
\centering
\caption{Best-fitting spectral parameters of 4U 1907+09 for different spectral models. The hydrogen column density ($N_H$) is fixed at $1.5\times10^{22}\;\mathrm{cm^{-2}}$. Errors are reported at 90\% confidence level.}
\begin{tabular}{lcccc}
\hline
Components & Parameters  &  M1 & M2 & M3 \\ \hline
\texttt{Const}      & $C_\mathrm{FPMB}$     & $1.003_{-0.006}^{+0.006}$ & $1.004_{-0.006}^{+0.006}$ & $1.004_{-0.005}^{+0.005}$ \\
\texttt{Powerlaw}   & $\Gamma$       & $1.12_{-0.10}^{+0.10}$ & - & - \\
    & $\mathrm{Norm}$       & $3.12_{-0.06}^{+0.06}\times10^{-2}$ & - & - \\
\texttt{highecut}      & $E_\mathrm{cutoff}$(keV)   &  $11.8_{-0.3}^{+0.3}$ & - & - \\
           & $E_\mathrm{fold}$ (keV)    &  $12.8_{-1.0}^{+1.8}$ & - & - \\
\texttt{compTT} & $T0$ (keV)       & - & $0.52_{-0.15}^{+0.07}$ & - \\
    & $kT_\mathrm{e}$ (keV)       & - & $7.9_{-1.4}^{+2.5}$ & - \\
    & $\mathrm{\tau}$  & - & $6.3_{-0.2}^{+0.3}$ & - \\
    & norm ($\times 10^{-2}$)      & - & $1.4_{-0.1}^{+0.2}$ & -  \\
\texttt{compmag}   
    & $kT_\mathrm{e}$ (keV)       & - & - & $3.4_{-0.1}^{+0.2}$  \\
    & $\mathrm{\tau}$  & - & - & $1.8_{-0.1}^{+0.2}$ \\
\texttt{bbodyrad}   & $kT_\mathrm{bb}$ (keV)   & - & - & $0.90_{-0.04}^{+0.03}$ \\
    & norm   & - & - & $18.5_{-2.6}^{+5.0}$  \\
\texttt{gabs1}       & $E_\mathrm{CRSF}$(keV)     &   $17.6_{-0.2}^{+0.1}$ & $17.9_{-0.1}^{+0.2}$ & $17.6_{-0.2}^{+0.3}$  \\
           & $\sigma_\mathrm{CRSF}$ (keV)&   $2.2_{-0.4}^{+0.5}$ &  $2.9_{-0.1}^{+0.2}$ &  $2.7_{-0.3}^{+0.4}$\\
           & strength    &    $1.8_{-0.4}^{+0.5}$ & $4.5_{-0.4}^{+0.5}$ & $3.1_{-0.5}^{+1.2}$  \\
\texttt{gabs2}       & $E_\mathrm{CRSF}$(keV)     &   $38.0_{-0.8}^{+0.9}$ & $39.6_{-1.3}^{+0.7}$ & $34.9_{-0.7}^{+0.9}$  \\
           & $\sigma_\mathrm{CRSF}$ (keV)&   $5.9_{-0.8}^{+1.0}$ &  $10.0_{-1.1}^{+1.3}$ &  $4.0_{-0.8}^{+1.7}$\\
           & strength    &    $28.4_{-6.0}^{+8.9}$ & $84.4_{-25.2}^{+8.9}$ & $11.2_{-2.9}^{+7.1}$  \\
\texttt{gauss}      & $E_\mathrm{Fe}$ (keV)       &    $6.31_{-0.06}^{+0.05}$ & $6.29_{-0.07}^{+0.06}$ & $6.18_{-0.07}^{+0.11}$ \\
           & $\sigma_\mathrm{Fe}$ (keV)   &    $0.11_{-0.11}^{+0.12}$ & $0.24_{-0.11}^{+0.13}$ & $0.64_{-0.15}^{+0.16}$\\
           & norm ($\times 10^{-4}$)       &    $1.8_{-0.5}^{+0.5}$ & $2.8_{-0.6}^{+0.7}$ & $7.9_{-0.2}^{+0.3}$ \\
           & EQW (eV) & 46 & 70 & 207 \\\hline
           & $\chi^2$/dof    &   1277/1110 & 1285/1110 & 1236/1110 \\\hline 
\end{tabular}
\\
\begin{minipage}{12cm}
\vspace{0.1cm}
\small  M1: \texttt{const*tbabs(gauss+highecut*powerlaw)*gabs*gabs}\\
M2: \texttt{const*tbabs(gauss+compTT)*gabs*gabs}\\
M3: \texttt{const*tbabs(gauss+bbodyrad+compmag)*gabs*gabs}

\end{minipage}
\label{table_par}
\end{table*}

\begin{table*}[]
\centering
\caption{Best-fitting spectral parameters for off-state, low flux on-state and high flux on-state. The hydrogen column density ($N_H$) is fixed at $1.5\times10^{22}\;\mathrm{cm^{-2}}$. The unit of flux is $(\times10^{-10})\;\mathrm{erg\;cm^{-2}\;s^{-1}}$. Errors are reported at 90\% confidence level.}
\begin{tabular}{lcccc}
\hline
Components & Parameters  & off-state & High flux on-state & Low-flux on-state \\ \hline
\texttt{Const}      & $C_\mathrm{FPMB}$     & $0.977_{-0.041}^{+0.043}$ & $0.996_{-0.008}^{+0.008}$ & $1.009_{-0.009}^{+0.009}$ \\
\texttt{Powerlaw}   & $\Gamma$       & $1.43_{-0.07}^{+0.07}$ & $1.12_{-0.02}^{+0.02}$ & $1.08_{-0.02}^{+0.02}$ \\
    & $\mathrm{Norm}$       & $4.34_{-0.53}^{+0.61}\times10^{-3}$ & $6.59_{-0.20}^{+0.21}\times10^{-2}$ & $1.84_{-0.05}^{+0.06}\times10^{-2}$ \\
\texttt{highecut}      & $E_\mathrm{cutoff}$(keV)   &  $22.9_{-2.5}^{+1.1}$ & $19.3_{-0.8}^{+0.5}$ & $18.3_{-0.5}^{+0.4}$ \\
           & $E_\mathrm{fold}$ (keV)    &  $4.6_{-1.5}^{+2.8}$ & $4.6_{-0.3}^{+0.4}$ & $6.3_{-0.4}^{+0.6}$  \\
\texttt{gabs1}       & $E_\mathrm{CRSF}$(keV)     &   $18.6_{-0.6}^{+0.6}$ & $18.3_{-0.4}^{+0.3}$ & $18.1_{-0.2}^{+0.2}$  \\
           & $\sigma_\mathrm{CRSF}$ (keV)&   $2.2_{-0.4}^{+0.5}$ &  $3.1_{-0.2}^{+0.2}$ &  $2.5_{-0.1}^{+0.2}$\\
           & strength    &    $3.4_{-1.0}^{+1.1}$ & $6.6_{-0.8}^{+0.7}$ & $5.1_{-0.4}^{+0.5}$  \\
\texttt{gabs2}       & $E_\mathrm{CRSF}$(keV)     &   - & - & $33.7_{-0.8}^{+1.0}$  \\
           & $\sigma_\mathrm{CRSF}$ (keV)&  - &  - &  $3.0_{-0.8}^{+1.3}$\\
           & strength    &    - & - & $5.7_{-2.0}^{+3.6}$  \\
\texttt{gauss}      & $E_\mathrm{Fe}$ (keV)       &    - & - & $6.26_{-0.08}^{+0.07}$ \\
           & $\sigma_\mathrm{Fe}$ (keV)   & - & - & $0.16_{-0.13}^{+0.11}$\\
           & norm ($\times 10^{-4}$)       &  - & - & $1.5_{-0.4}^{+0.5}$ \\\hline
           & flux$_{3-25\;\mathrm{keV}}$  & $0.47_{-0.02}^{+0.02}$ & $12.16_{-0.08}^{+0.08}$ &  $3.91_{-0.03}^{+0.03}$ \\\hline
           & $\chi^2$/dof    &   205/181 & 1045/929 & 1047/940 \\\hline 
\end{tabular}
\\
\label{table_flux}
\end{table*}

\section{Discussion} \label{sec:discussion}
4U 1907+09 is a classical sgXB known to show dips and flares in its light curve. In this work, we analyzed the \textit{NuSTAR} data of 4U 1907+09 observed during November 2024. 
The spin down/spin up and torque reversals of the pulsar were documented in several studies \citep{Cook1987, intzand1998, Baykal2001, Baykal2006, Mukerjee2001, Fritz2006, Inam2009, Sahiner2012, Varun2019, Tobrej2023}. In the most recent study, \cite{Tobrej2023} found that 4U 1907+09 has been spinning down. In this work, we estimated the pulse period of 4U 1907+09 to be $\sim443.99(4)$ s. This indicates that the source is still spinning down. 
At low or moderate luminosities, a hot quasi-static shell develops as captured wind matter settles subsonically on the NS magnetosphere \citep{Davies1981}. 
During spin-down episodes, this shell removes angular momentum from the rotating NS magnetosphere through large-scale convective processes \cite{Shakura2012} and could be responsible for 4U 1907+09's long-term spin-down behaviour.

The pulse profile is one of the indicators of the emission region from the neutron star. A double-peaked (at $\sim0.26$ and $0.73$) asymmetric pulse shape was observed in 4U 1907+09, as shown in the left panel of Figure \ref{pulse}. A similar pulse shape was also reported for the 4U 1907+09 in the previous studies \citep{Maitra2013, Varun2019, Tobrej2023}. The observed asymmetric pulse shape is most likely caused by a magnetic dipole offset from the rotational axis of the pulsar \citep{Parmar1989, Leahy1991, Riffert1993, Bulik1995}. Another plausible explanation for the asymmetric shape of pulse profiles is an asymmetric accretion stream \citep{Basko1976, Wang1981, Miller1996}. As energy increases, the pulse shape changes significantly. The peak at 0.26 begins to fade at 15 keV. For Vela X-1, \cite{Shakura2013} revealed that when the source flips from on-state to off-state, it transitions from a fan beam pattern to a pencil beam pattern in the 20–60 keV energy range. Furthermore, Vela X-1 exhibited absorption in the pencil beam at lower energies.
However, 4U 1907+09 displayed the pencil beam during the on-state, without indicating a fan beam pattern. Furthermore, no absorption was detected in the pencil beam at low energies. It indicates that the optical depth of accretion flow above the polar cap is always less than 1 in 4U 1907+09 during the transition from on-state to off-state. The PF of X-ray pulsars shows the complex behaviour and dips at cyclotron energy \citep{Lutovinov2009, Ferrigno2023}. \cite{Tobrej2023} observed a dip near the cyclotron energy in PF. The PF in our investigation showed increasing behaviour with energy and did not exhibit a complicated structure because of the limited number of data points.

The \textit{NuSTAR} spectrum was fitted with three different models: power law with \texttt{highecut}, \texttt{compTT}, and \texttt{compmag}. Table \ref{table_par} contains the corresponding fit parameters. The iron line and CRSF with its harmonics were found at $\sim6.31$ keV, $\sim17.6$ keV, and $\sim38.02$ keV, respectively. After incorporating gravitational redshift, the CRSF corresponds to a magnetic field of $\sim2\times10^{12}$ G. The average flux of the source during the on-state in the 3–50 keV energy range was found to be $\sim 6.59\times10^{-10} \mathrm{erg\;cm^{−2}\;s^{−1}}$. Assuming a distance of 1.9 kpc, the estimated luminosity of the source is $\sim2.85 \times 10^{35} \mathrm{erg\;s^{−1}}$.
The luminosity of the source is similar to that observed by \textit{NuSTAR} in 2018 \citep{Tobrej2023}. Using the formalism in \cite{Becker2012}, we estimate the critical luminosity $L_{\mathrm{crit}} = 1.5\times10^{37}\;\left(\frac{B}{10^{12}\;\mathrm{G}}\right)^{16/15}\;\mathrm{erg\;s^{-1}}$ to be $3\times10^{37}\mathrm{erg\;s^{-1}}$. The Coulomb braking luminosity $L_{\mathrm{coul}} = 1.2\times10^{37}\;\left(\frac{B}{10^{12}\;\mathrm{G}}\right)^{1/3}\;\mathrm{erg\;s^{-1}}$ to be $1.5\times10^{37}\mathrm{erg\;s^{-1}}$ \citep{Becker2012, Langer1982}. 
The geometry of the accretion column and emission beam pattern changes with the luminosity. At high luminosity
($L_X \sim 10^{37−38}\;\mathrm{erg\;s^{−1}}$), the radiation predominantly escapes through the column walls in the sinking zone, resulting in a "fan beam" \citep{Davidson1973}. At low luminosity ($L_X \sim 10^{35}\;\mathrm{erg\;s^{−1}}$), 
the infalling matter impacts the NS surface, and radiation is released from the top of the column, resulting in a "pencil beam" pattern \citep{Burnard1991, Nelson1993}. In the intermediate range, $L_X \sim 10^{35−37}\;\mathrm{erg\;s^{−1}}$, 
the radiation emits from the wall as well as the top of the accretion column, resulting in a complicated emission pattern composed of fan beams and pencil beams \citep{Blum2000}.
The observed luminosity of 4U 1907+09 is much lower than the $L_\mathrm{crit}$ and $L_\mathrm{coul}$. As a result, radiation from a gas-mediated shock without collision is anticipated. This leads to a "pencil" beam, where the radiation is emitted mostly along the magnetic field lines.

Numerous studies examine the dips and flares in the light curve of 4U 1907+09 \citep{Intzand1997, Doroshenko2012, Sahiner2012, Shakura2012, Varun2019}. Apart from 4U 1907+09, off-states or dips have been reported in the Vela X-1 \citep{Kreykenbohm2008, Doroshenko2011} and GX 301-2 \citep{Gogus2011}. Previous studies have ruled out the likelihood of dips generated by a dense, clumpy wind obscuring the neutron star \citep{Furst2011, Sahiner2012}. \cite{Shakura2012} suggested that the onset of the off-state in these systems is due to the transition from the Compton cooling-dominated to the radiative cooling-dominated regime. During this observation, the source showed a dip and flare in the light curve. During the dip, the flux of the source decreased 8 times the flux during the low-flux on-state. The flux during high flux on-state increased 3 times during low flux on-state. As the flux varied during the on-state, the pulse profile did not differ much and followed a similar trend (as shown in the right panel of Figure \ref{pulse}). This indicates that the emission process did not change during the flare activity. 
The cyclotron energy correlates with the luminosity \citep{Staubert2019}. In the super-critical state, the cyclotron energy displays a negative correlation with the luminosity. The cyclotron energy remains constant or has a positive correlation with luminosity in the sub-critical phase.
The spectral modelling of 4U 1907+09 revealed that the $E_\mathrm{CRSF}$ is unaltered at $\sim18$ keV in both the high flux (flare) on-state and the off-state. It suggests that the accretion column may not be formed. The off-state spectrum is softer than the on-state spectrum. This behaviour is also seen in GX 301-2 \citep{Gogus2011} and Vela X-1 \citep{Doroshenko2011}. The e-folding energy during the high flux on-state is lower than the low flux high state. 
The line width of the CRSF during the high flux on-state is higher than the low flux on-state.

In addition, the polarimetric studies of strongly magnetized X-ray pulsars are useful for the determination of XRP geometry and related astrophysical phenomena \citep{Meszaros1988, Santangelo2019, Caiazzo2021, Poutanen2024}. 4U 1907+09 is an ideal candidate to study the polarimetric properties in the subcritical accretion regime. For 4U 1907+09, the vacuum polarization effects are anticipated to be important \citep{Shakura2013}.
For Vela X-1 (a wind-fed sgXB similar to 4U 1907+09), the Imaging X-ray Polarimetry Explorer (IXPE; \cite{Weisskopf2022}) found a polarization degree (PD) of 2.3\% ± 0.4\% at the polarization angle (PA) of −47.3 ± 5.4 \citep{Forsblom2023}. Finally, we can speculate a small PD for 4U 1907+09 as detected in Vela X-1.

\section{Summary} \label{sec:summary}
The results and findings of this work are summarized as follows:
\begin{itemize}
  \setlength{\itemsep}{0pt}
  \setlength{\parskip}{0pt}
  \setlength{\parsep}{0pt}
    \item 4U 1907+09 exhibiting normal flux, flare, and dip (off-state) in the light curve and providing insights into the variability of the source.
    \item With an estimated pulse period of $443.99(4)$ s, the source was determined to be spin down. 
    \item During the on-state, the typical spectral and temporal characteristics of 4U 1907+09 are detected, including a double-peaked pulse profile, energy dependence of pulse profile, CRSF, and dips in the light curve.
    \item The CRSF at $\sim18$ keV was found in both on and off states. While the CRSF energy remained constant, CRSF's width showed a positive correlation with flux.
    \item  The spin phase influenced the spectral properties. Cutoff energy is in phase with the pulse shape, but photon index and e-fold energy are out of phase.
\end{itemize}

\section*{Acknowledgments}
We thank the anonymous referee for their comments and recommendations, which enhanced the quality of this paper. We have utilized the archived \textit{NuSTAR} data provided by the High Energy Astrophysics Science Archive Research Center (HEASARC)
online service maintained by the Goddard Space Flight Center. For this work, we have made use of the \textit{NuSTAR} Data Analysis Software
(NuSTARDAS) jointly developed by the ASI Space Science Data Center (SSDC, Italy) and the California Institute of Technology
(Caltech, USA). RK acknowledges the Department of Astronomy and Astrophysics, Tata Institute of Fundamental Research, Mumbai, India, for providing research facilities.

  \bibliographystyle{elsarticle-harv} 
  \bibliography{bib}

\end{document}